\documentclass[aps,prl,twocolumn,groupedaddress,showpacs]{revtex4}
\usepackage{amsmath}
\usepackage{amssymb}
\usepackage{bm} 
\usepackage{graphicx}
\usepackage[latin1]{inputenc}

\newcommand{\myscaleboxB}[1]{\scalebox{0.42}[0.42]{#1}}
\newcommand{\be}{\begin{equation}}
\newcommand{\ee}{\end{equation}}

\begin{document}

\title{Droplets in the two-dimensional $\pm J$ spin glass: evidence
  for (non-) universality}

\author{A. K. Hartmann}
\affiliation{Institut f\"ur Physik, Universit\"at Oldenburg,
 26111 Oldenburg, Germany}

\begin{abstract}
Using mappings to computer-science problems and by
applying sophisticated algorithms, one can study numerically 
many problems much better compared to applying standard approaches
like Monte Carlo simulations. Here,
using calculations of ground states of suitable perturbed systems,
droplets are obtained in  two-dimensional $\pm J$ spin glasses, which
are in the focus of a currently very lifely debate. 
Since a sophisticated matching algorithm is applied here, exact
ground states of large systems up to $L^2=256^2$ spins can be
generated. Furthermore, 
no equilibration or extrapolation to $T=0$ is necessary.
Three different $\pm J$ models are studied here: a) with open boundary
conditions, b) with fixed boundary conditions and c) a diluted system
where a fraction $p=0.125$ of all bonds is zero.
For large systems, the droplet energy shows for all three models a
power-law behavior $E_{\rm D}\sim L^{\theta^\prime_{\rm D}}$ with
$\theta^\prime_{\rm D}<0$. This is different from
previous studies of domain walls, where a convergence to a constant
non-zero value ($\theta_{\rm dw}=0$) 
has been found for such  models. After correcting
for the non-compactness of the droplets, the results are likely 
to be compatible
with  $\theta_{\rm D}\approx -0.29$ for all three models.
 This is in accordance with
the Gaussian system where $\theta_{\rm D}=-0.287(4)$ 
($\nu\approx3.5$ via $\nu=-1/\theta_{\rm D}$). 
Nevertheless, the disorder-averaged 
spin-spin correlation exponent $\eta$ 
is determined  here via the probability
to have a non-zero-energy droplet, and $\eta\approx0.22$ is
found for all three models, this being in contrast to the behavior of
the model with Gaussian interactions, where exactly $\eta=0$.
\end{abstract}
\maketitle

Monte Carlo simulations \cite{newman1999,landau2000}
and related approaches are  common
ways to study physical systems numerically, in particular for the usual case
where no exact analytic solution can be provided.
So far, it has only occasionally been recognized that by using mappings to
computer-science problems and by applying sophisticated algorithms
one can obtain in many cases results which are superior in
comparison to using standard algorithms. 

Here, Ising spin glasses are considered, which are the 
most-frequently studied systems in
statistical physics \cite{binder1986,mezard1987,fischer1991,young1998}. 
However, despite more than two
decades of intensive research, many properties of spin glasses,
especially in finite dimensions, are still not well understood. 
For two-dimensional spin glasses it is now widely accepted that no
ordered phase for finite temperatures exists 
\cite{rieger1996,kawashima1997,carter2002,stiff2d} in
this case. Nevertheless, the $\pm J$ model with a bimodal distribution of the
interactions is in the center of a currently very lively debate
\cite{houdayer2001,stiff2d,amoruso2003,%
lukic2004,poulter2005,wang2005,katzgraber2005,%
fisch2006a,joerg2006,fisch2006b,fisch2006,katzgraber2007}, 
in particular whether the behavior is equivalent to that of
the model with a Gaussian distribution of the interactions. One
central question is whether the correlation length, when approaching
$T_c=0$, diverges algebraically \cite{wang1988} 
$\sim T^{-\nu}$, as for the Gaussian
model,  or like an exponential \cite{saul1994} $\sim
T^{-2}e^{-C/T}$, formally equivalent to $\nu=\infty$. 
In particular, Ref.\ \onlinecite{joerg2006} claims evidence for a power-law
divergence with the same critical exponent $\nu\approx 3.5$ as the
Gaussian system, and that the
spin-spin correlation exponent $\eta=0$. This exponent is defined via
$[\langle S_i S_{i+l} \rangle^2]_J
\sim l^{-\eta}$, $[\ldots]_J$ and $\langle\ldots\rangle$
being the averages over the quenched
disorder and the thermal average, respectively. 

Unfortunately, most of the above cited work is based on
finite-temperature calculations, in particular Monte Carlo
simulations, hence an extrapolation $T\to T_c=0$ 
is necessary. Also the systems are restricted in most cases to
rather small sizes $L\le 64$. 
Only in the case, where parallel tempering
Monte Carlo simulations \cite{katzgraber2005,katzgraber2007} 
or the worm algorithm \cite{wang2005} have been used, 
sizes $L\le 128$ could be considered. Consequently, 
in Ref.\ \onlinecite{katzgraber2007} it
has been shown that the presently available finite-temperature data
does not allow to draw final conclusions.
In Ref.\ \onlinecite{poulter2005} also exact $T=0$ properties are
calculated (up to $L=128$ for $L\times L$ systems and up to $L=64$ for
$9L\times L$ systems), but only the spin-spin correlation could be obtained by
this approach.

Here, we go much beyond the previous work.
Exact ground-state (GS) calculations \cite{opt-phys2001}
are applied, which allows to obtain GSs
\cite{stiff2d,aspect-ratio2002,droplets2003,droplets_long2004} 
for large systems like 
$L=480$. The method relies on mapping the GS
calculation to a graph-theoretical problem and using sophisticated
algorithms developed in computer science. 
By using suitable perturbations of the original systems,
one can go beyond pure GS calculations and study excitations
like domain walls (DWs) \cite{stiff2d,aspect-ratio2002} and droplets 
\cite{droplets2003,droplets_long2004}. Hence, large systems in exact 
equilibrium can be investigated and no extrapolation to $T=0$ is necessary.

In Refs.\
\cite{stiff2d,aspect-ratio2002,droplets2003,droplets_long2004} 
this approach has been used to show
that for the two-dimensional spin glass with Gaussian disorder all
assumptions made by the droplet 
theory \cite{mcmillan1984,bray1987,fisher1986,fisher1988} are fulfilled.
In particular the energy scaling of the basic excitations 
DWs and droplets follows
power laws $E_{\rm dw}\sim L^{\theta_{\rm dw}}$ and 
$E_{D}\sim L^{\theta_D}$, respectively, with the same universal value
$\theta \equiv\theta_{\rm D}=\theta_{\rm dw}$, which is related \cite{bray1984} 
to the correlation-length exponent via
$\theta=-1/\nu = -0.287(4)$. 
 On the other hand,
for the $\pm J$ model, the average
energy of DWs approaches a constant for large system sizes
\cite{stiff2d} $L\ge 128$, i.e. $\theta_{\rm dw}=0$. 
This appears to be, via $\theta=1/\nu$, compatible with an exponential
divergence of the correlation length, if either the low-temperature
behavior is dominated by DWs, or if the scaling of the energy of
droplets, which are expected to be the dominant excitations, is the
same as for DWs.

In this work, droplet excitations for three different models with
discrete distributions of the interactions 
are calculated using an approach
which is based on
exact GS calculations.  This approach allows 
to consider large systems up to $L=256$, no extrapolation of the temperature
and no equilibration are
necessary.  The
main result is that the scaling behavior is different from the scaling
of DWs, i.e. $\theta_{\rm D}<0$. This is compatible with a
power-law divergence of the correlation length, as recently claimed
\cite{joerg2006}. After correcting for the non-compactness of the
droplets \cite{kawashima2000}, the results seem to be in all three
cases the same as for
the Gaussian model, hence Gaussian and discrete models appear to be in
the same universality class regarding the low-temperature behavior. 
Furthermore, in this work the scaling
of the spin-spin correlation function is studied, resulting in
$\eta\approx 0.22$ for all three models.

The Hamiltonian which is studied here is the usual  Ising
spin glass model:
\begin{equation}
{\mathcal H}=-\sum_{\langle i,j\rangle }J_{ij}S_iS_j,
\end{equation}
where the spins $S_i=\pm1$ 
lie on the sites of a square lattice with $N=L^2$ sites,
 the bonds $J_{ij}$ couple nearest-neighbor sites on the lattice.
$J_{ij}= \pm J$ with equal probability for the quenched realizations
of the disorder. Here systems with either open boundary conditions
(bc) in all directions, for the initial GS calculation, or fixed bc always
are studied. For the latter case, also diluted
samples are considered, where each bond is set to zero with
probability $p=0.125$;

Here, droplets are considered, 
as introduced by Kawashima \cite{kawashima2000}, which are lowest-energy
excitations with respect to the GS. 
They consist of a connected cluster of spins, 
which include a certain pre-selected spin, here a center spin of
the system. The spins at the boundary are fixed to their GS
orientations. The energy scaling of these droplets is expected to follow
a $L^{\theta^\prime_{\rm D}}$ scaling, where \cite{kawashima2000}
$\theta^\prime_{\rm D}=\theta_{\rm D}D_V/D$ is related to the volume fractal
dimension $D_V$ of the droplets, the dimension $D=2$ of the system and
the usual droplet exponent $\theta_{\rm D}$.
 The main approach used here is based on
mapping \cite{bieche1980,barahona1982b,derigs1991} 
the GS
calculation to the {\em minimum-weight perfect matching problem} and
using sophisticated matching algorithms from graph theory. For details,
please see the pedagogical description in
Ref.\ \onlinecite{opt-phys2001}. The droplet calculation of each 
disorder realization is
based on a sequence of $2L$ suitable modifications of the disorder,
each time followed by a GS calculation. The details of the
algorithm \cite{meandering_short} 
are described extensively in  Refs.\
\cite{droplets2003,droplets_long2004}.

Minimum-energy droplets have been obtained using the aforementioned
approach for system sizes $L=6$ to $L=256$ ($L=160$ for the fixed
bc). All results are averages
over many disorder realizations, the number of realizations is between
 20000 for small sizes and 5000 for the largest sizes.

\begin{figure}[htb]
\begin{center}
\myscaleboxB{\includegraphics{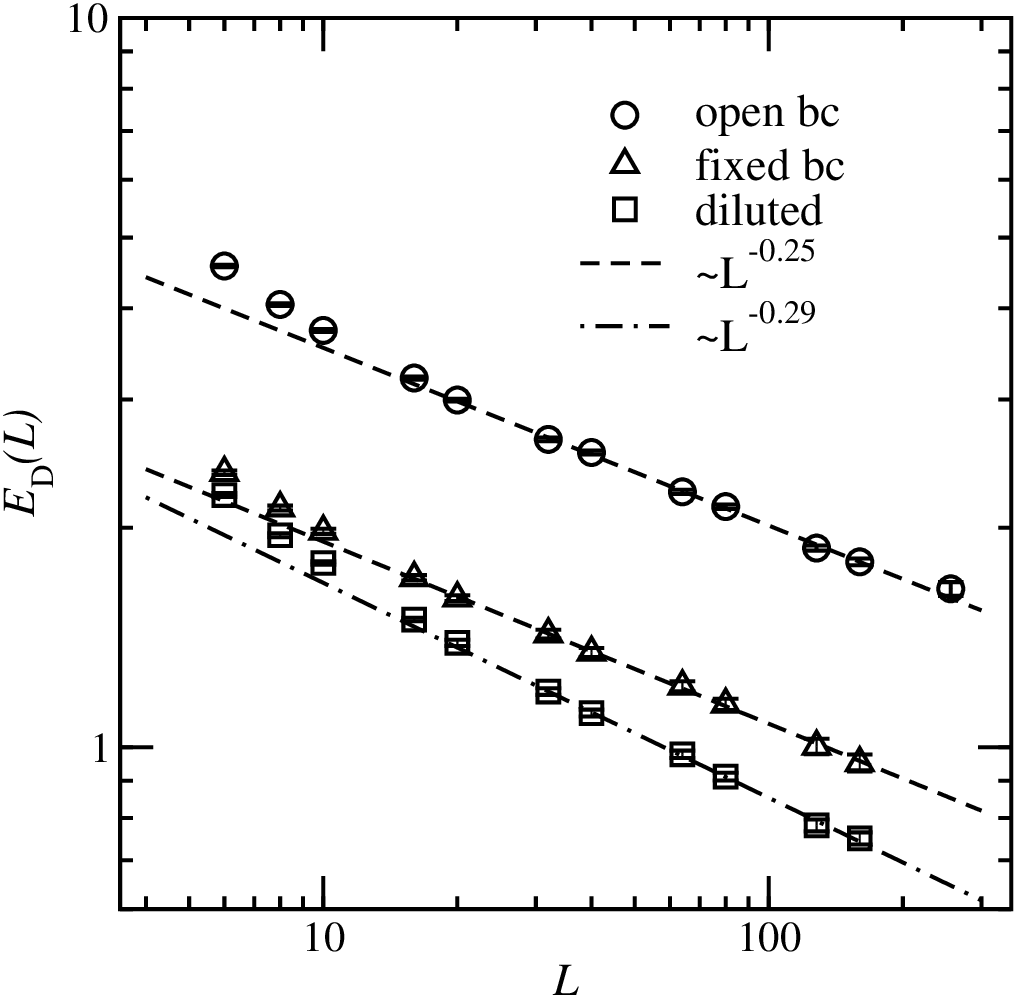}}
\end{center}
\caption{Average droplet energy $E_{\rm D}$ as a function of system
  size in a double logarithmic plot. %
The solid line sows a fit to a power law
  with correction to scaling, while the dashed lines show the results of
fits to a simple power laws.}
\label{fig:EL}
\end{figure}

In Fig.\ \ref{fig:EL} the average droplet energy $E_{\rm D}$ is shown as a
function of system size $L$. For large system sizes, a power-law behavior
is visible. This 
 corresponds \cite{bray1984} to  a power-law divergence of the correlation 
length as observed recently  by J\"org et al \cite{joerg2006}, as in
contrast to the behavior of DWs for the $\pm J$ model and other models
exhibiting a quantized energy spectrum.
 A fit to
the function $AL^{\theta^\prime_{\rm D}}$ for $L\ge 32$ yields 
$\theta^\prime_{\rm  D}=-0.244(6)$ for open bc with 
a good quality of the fit \cite{quality,press1995}
$Q=0.32$. 
Similarly, $\theta^\prime_{\rm D}=-0.250(6)$ ($Q=0.82$) are
obtained for fixed bc and   $\theta^\prime_{\rm D}=-0.295(6)$
($Q=0.64$) for the diluted systems.

\begin{figure}[htb]
\begin{center}
\myscaleboxB{\includegraphics{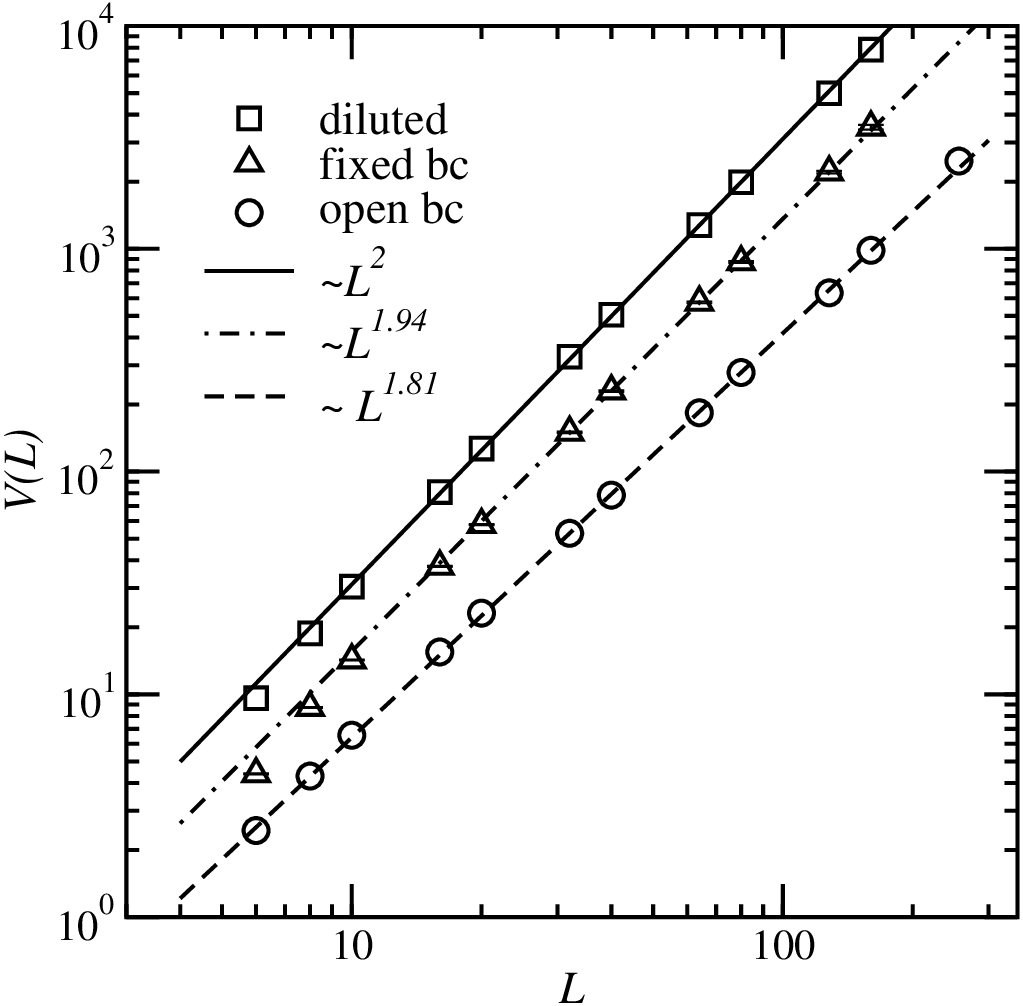}}
\end{center}
\caption{Volume $V$ of the droplets as a function of
  the system size, respectively. Error bars are smaller than symbol
  sizes. The lines show the results of
  power-law fits of $\sim L^{D_V}$ to $V(L)$ with
$D_V=1.81$, $D_V=1.94$ and $D_v=2$ for open bc, fixed bc, and the
diluted system, respectively. The data for the diluted system has been
shifted up by a factor of 2, for better visibility.}
\label{fig:SVL}
\end{figure}

To obtain the droplet exponent $\theta$, 
the geometric properties of the droplet volumes are considered next. In
Fig.\ \ref{fig:SVL} the volume $V$ of the droplets
is shown as a function of the system size $L$ for the three different
models. Note that due to the degeneracy of the models, droplets with
many different values for the volume are possible. Unfortunately,
unless through complete enumeration, no algorithm to sample GSs or
droplets with the same weight/probability is known. Here, the
degeneracy is broken, by selecting the droplets with the smallest {\em
surface} \cite{smallestDW2007}, hence the behavior of the volume is
not controllable and can give only a rough idea of the true behavior.
When fitting a power
law $\sim L^{D_V}$ to $V(L)$ of the open bc model, 
one obtains $D_V=1.81(1)$,
$D_V=1.94$ for fixed bc, while the behavior of the diluted system is
compatible with $D_V=2$. Via considering
 $\theta_{\rm D}=\theta^\prime_{\rm D}D/D_V$, it
appears likely that $\theta_{\rm D}\approx-0.29$ universally for all
three models. Since the low-temperature behavior is dominated by
droplets, and not by DWs, this would mean the $\pm J$ models not only show
a power-law behavior for the divergence of the correlation
length like the Gaussian system, but exhibit
the {\em same}  value $\nu=-1/\theta_{\rm D}\approx 3.5$ for the
correlation-length exponent. 
In any case, even if $\nu$ could still be slight different for Gaussian and
$\pm J$ Ising spin glasses,  the results show unambiguously 
that the scaling behavior of
droplets and DWs is different in the $\pm J$ case, the main reason
\cite{amoruso2003}
being the discreteness of the spectrum of excitations.
This is opposed to the assumptions of the droplet 
theory \cite{mcmillan1984,bray1987,fisher1986,fisher1988}, were the behavior
of all types of excitations is governed by one single exponent 
$\theta=\theta_{\rm D}=\theta_{\rm dw}$.


\begin{figure}[htb]
\begin{center}
\myscaleboxB{\includegraphics{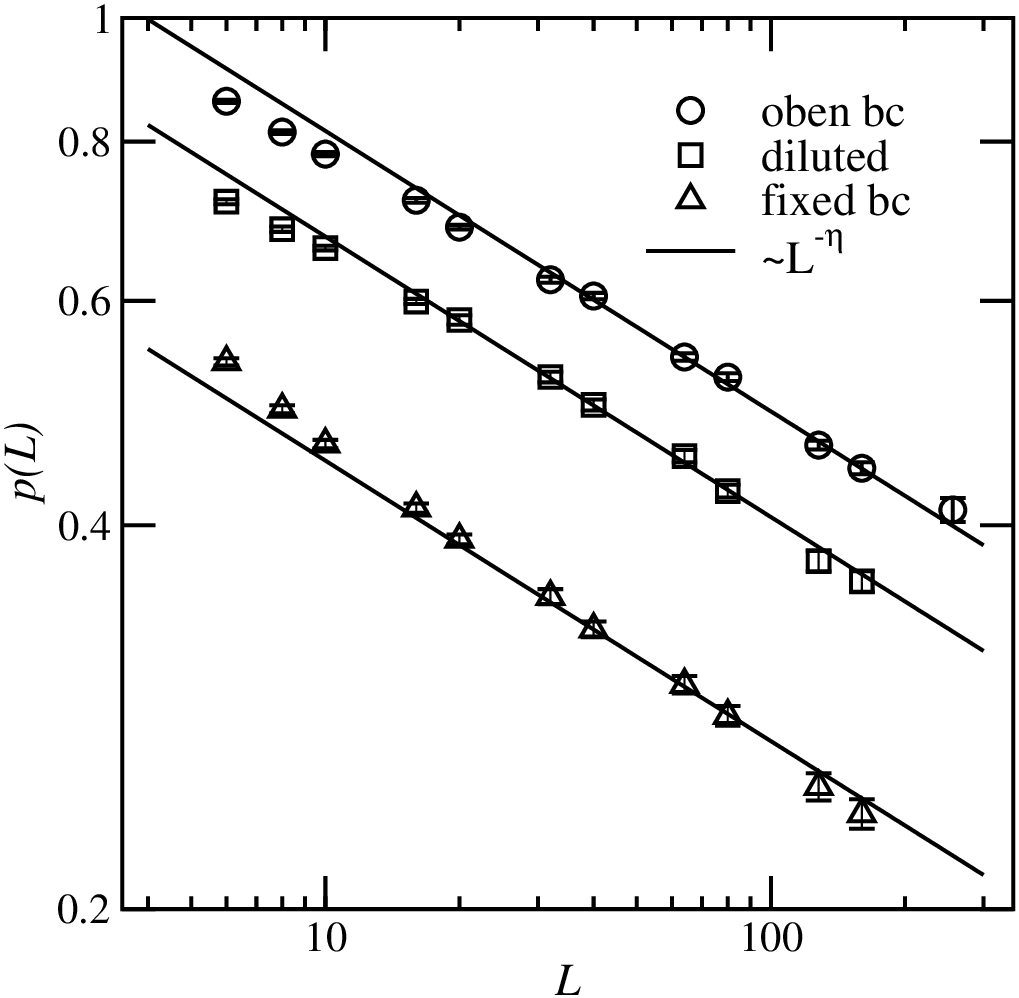}}
\end{center}
\caption{Probability $p(L)$ that a droplet has non-zero energy
as a function of system
  size in a double logarithmic plot, for open bc, fixed bc and the
  diluted system, respectively.
The lines show simple power laws $\sim L^{-\eta}$, with $\eta=0.22$ in
all three cases.}
\label{fig:PE0L}
\end{figure}

Finally, to determine $\eta$, here the relation \cite{bray1987} $[\langle S_i
S_{i+l} \rangle^2]_J = p(L)$ is used, where $p(L)$ is the probability
to have a droplet with non-zero energy. 
In Fig.\ \ref{fig:PE0L} $p(L)$ is shown as a function of system
size. According droplet theory \cite{bray1987}
one expects $p(L)\sim L^{-\eta}$. When fitting  for
system sizes $L>32$, we obtain in all three cases 
$\eta$ close to $=0.22$ with good qualities of the fit, showing the
universality of this result. This makes it likely that also the behavior
of the average droplet energy is universal, hence the 
differences  observed in Fig.\ \ref{fig:EL} are indeed through
different fractal dimensions $D_V$.
Note that if one related
$\eta$ to the probability that a non-zero-energy cross-system DW
exists, one would assume $\eta=0$ since the DW energy settles
at a constant value for large systems \cite{stiff2d},
i.e. $\theta_{\rm dw}=0$. Nevertheless, two spins are uncorrelated if
there exists {\em any} zero-energy excitation separating them, hence one has
to consider droplets as well, as performed here. 

The result is compatible with $\eta=0.21$ obtained \cite{ozeki1990} 
from $T=0$ transfer-matrix calculations of the correlation function
for $L\le 12$. On the other hand,
the claim $\eta=0$ of Ref.\
\onlinecite{joerg2006}, which is based  on an  
extrapolation $T\to 0$ for small systems $L\le 64$,x2
 is clearly ruled out.  
Even more, fits of the actual data obtained in Ref.\
\onlinecite{joerg2006} yielded, depending on the fits, also values $\eta>0$.
In Ref.\ \onlinecite{katzgraber2005} a value $\eta=0.138$ was found
via Monte-Carlo simulations at finite but low temperature for system
sizes $L\le 128$,
but it was mentioned by the authors that their results are compatible with
a ``large range of $\eta$ values''. 
Interestingly, at higher temperatures, a higher effective
exponent $\eta_{\rm eff}\approx 0.2$ was observed by the authors.
A value $\eta=0.14(1)$ has been obtained in Ref.\
\onlinecite{poulter2005}, but that result is obtained from studying only four
different system sizes, and the result depends on the 
assumption that the
correlation length at $T=0$ diverges as $L^{3/2}$. 
Note that for the Gaussian model at exactly $T=0$, due to the
uniqueness of the GS, we have 
$\eta=0$, hence regarding this quantity, the two classes of
models look non-universal.

One can understand by a simple scaling argument, why DWs and
droplets can behave differently in the $\pm J$ model. We denote by
$q_l$ the probability that a zero-energy DW exists in a system of
size $l$. Since this probability approaches \cite{stiff2d} a 
finite value $\tilde q$ for about  $L\ge 100$, we can assume
$q_l\approx \tilde q$ for simplicity. We look at the system at
different scales $l=l_0^0,l_0^1,l_0^2,\ldots,l_0^k=L$ ($l_0>1$
arbitrary, $k=\ln L /\ln l_0$), where we can assume the different
DWs are independent.
A non-zero droplet exists only if
 on {\em all} scales no zero-energy closed DW exists,
i.e. $p(L)=\prod_{i=0,\ldots,k} q_{l_0^i} \approx \tilde q^k = L^{-\eta}$
with $-\eta = \ln \tilde q / \ln l_0<0$. Hence, the probability for
non-zero energy droplets decreases with a power-law (hence the mean
droplet energy), while the probability for non-zero energy DWs
(and the mean DW energy) saturates for $L\to\infty$.

To summarize, droplet excitations for three different variants of the 
two-dimensional $\pm J$ Ising spin
glasses were studied. Here an advanced methodology from
graph theory is used, based on mapping the
GS calculation to the minimum-weight perfect matching
problem, using sophisticated matching algorithms from computer science and
studying sequences of suitable modified realizations of the disorder. 
This allows to treat at $T=0$, without need for an extrapolation of
the temperature, large systems up to $L=256$ exactly. The average
droplet energy shows a clear power-law behavior with exponent
$\theta^\prime_{\rm D}<0$. It appears likely 
that when taking the non-compactness of the droplets
into account, the same droplet exponent $\theta_{\rm D}\approx-0.29$
emerges for all three models studied here,
and hence the same value $\nu=-1/\theta_{\rm D}\approx3.5$ as
for the model with Gaussian disorder is obtained. On the other hand,
 the value $\eta\approx0.22$ for the exponent
describing the decay of the spin-spin correlations is obtained for all
three models, but this is clearly different from the Gaussian model, 
where $\eta=0$. Remarkably,
$\theta_{\rm D}\neq \theta_{\rm dw}=0$ can be explained by a
simple scaling argument.

\begin{acknowledgments}
The author thanks Ian Campbell,  Ron
Fisch, Helmut Katzgraber, Oliver Melchert, Mike Moore, Martin Weigel
and Peter Young for interesting discussions. He is grateful to
Oliver Melchert, Mike Moore and Peter Young 
for critically reading the manuscript.
Financial support was obtained from the
{\em VolkswagenStiftung} (Germany) within the program
``Nachwuchsgruppen an Universit\"aten'' 
and from the European Community via the DYGLAGEMEM contract.
\end{acknowledgments}

\bibliography{alex_refs,remarks}

\begin{thebibliography}{41}
\bibitem[{qua()}]{quality}
\bibinfo{note}{The quality $Q$ of the fit states the probability
  \cite{press1995} that the data matches to the fitting function as present or
  worse, under the assumption that the fitted function is indeed the correct
  model for the data.}
\end{thebibliography}

\end{document}